\documentclass{sig-alternate}
\usepackage{mathptmx} 
\usepackage{graphicx}
\usepackage{fancyhdr}
\usepackage[normalem]{ulem}
\usepackage[hyphens]{url}
\usepackage[sort,nocompress]{cite}
\usepackage[final]{microtype}
\usepackage[keeplastbox]{flushend}
\usepackage[bookmarks=true,breaklinks=true,colorlinks,citecolor=blue,linkcolor=blue,urlcolor=blue]{hyperref}
\usepackage{listings}       
\usepackage{xcolor}
\definecolor{commentgreen}{RGB}{2,112,10}
\usepackage{pifont}         
\usepackage{booktabs}       
\usepackage{authblk}        
\providecommand{\keywords}[1]{\textbf{\textit{\\Keywords -}} #1}

\fancypagestyle{firstpage}{
  \fancyhf{}
  
  \fancyfoot[C]{\thepage}
}

\author[]{Akhil Shekar}
\author[]{Morteza Baradaran}
\author[]{Sabiha Tajdari}
\author[]{Kevin Skadron\vspace{-1em}}
\affil[]{University of Virginia, School of Engineering and Applied Sciences, Charlottesville, VA, USA\vspace{-1em}}
\affil[]{\{as8hu, rgq5aw, jvx2tt, skadron\}@virginia.edu}
\title{HashMem : PIM-based Hashmap Accelerator\vspace{-2.5em}} 

\begin{document}
\maketitle
\thispagestyle{firstpage}
\pagestyle{plain}


\begin{abstract}

\textbf{Hashmaps are widely utilized data structures in many applications to perform a probe on key-value pairs. However, their performance tends to degrade with the increase in the dataset size, which leads to expensive off-chip memory accesses to perform bucket traversals associated with hash collision. In this work, we propose HashMem, a processing-in-memory (PIM) architecture designed to perform bucket traversals along the row buffers at the subarray level. Due to the inherent parallelism achieved with many concurrent subarray accesses and the massive bandwidth available within DRAM, the execution time related to bucket traversals is significantly reduced. We have evaluated two versions of HashMem, performance-optimized and area-optimized, which have a speedup of $49.1x/17.1x$ and $9.2x/3.2x$ over standard C$++$ map and hyper-optimized hopscotch map implementations respectively.}
\end{abstract}

\keywords{Processing In Memory, HashMaps, In-Situ Computing, DRAM, Memory Systems}

\section{Introduction}

As we move further into the digital age, the amount of data being generated and consumed daily is increasing at an unprecedented rate. One of the popular data structures for searching large datasets is a hashmap, due to its near-constant-time lookup performance. For large datasets, the hashmaps usually are not cache-resident, and hence, any lookup would be an expensive off-chip DRAM access to read the entire hash chain to perform a single probe. 

By supporting hashmap lookup directly in memory, i.e. using a processing-in-memory (PIM) architecture, applications can avoid costly memory transfers between the processor and memory, leading to faster and more efficient lookups. Hashmaps are particularly well-suited to PIM architectures, as they are able to leverage the parallel processing capabilities of the memory to perform lookups and hash-chain traversals in constant time.

For a brief background, a hashmap is a data structure that allows efficient storage and retrieval of key-value pairs. It uses a hash function to compute an index into an array of buckets or slots from which the desired value can be found. In other words, a hashmap is an associative array that maps keys to values.

The computation of hashmaps is memory-bound because it heavily relies on accessing memory when the dataset cannot fit within the cache. When searching for a key in a hashmap, the hash function is used to compute the index of the bucket where the key is stored. The bucket is then accessed in memory to retrieve the value associated with the key. As the size of the hashmap increases, the number of memory accesses required to perform a lookup also increases, making it a memory-bound operation.

Processing-in-memory is a new architecture paradigm that breaks down the memory wall by integrating processing elements within the memory chips themselves.  
This eliminates the need for data to be transferred between memory and processor, leading to significant improvements in performance and energy efficiency \cite{mutlu2022modern}.

In this paper, we propose HashMem, a PIM architecture designed to accelerate key-value probes on hashmaps. Our architecture comprises two versions: area-optimized and performance-optimized. Both place processing elements adjacent to each subarray in the DRAM, but the area-optimized provides one processing unit per subarray and operates on one value at a time, i.e., element-serial, bit-parallel; while the performance-optimized provides multiple processing units per subarray, operating on the entire row at once in an element-parallel but bit-serial fashion. Thus, in the latter case, the values are laid out in a column-oriented fashion, so that each row contains a single-bit slice from thousands of values, achieving high parallelism at the expense of requiring $b$ steps in order to find a b-bit key.

These organizations are evaluated against standard C$++$ map and hyper-optimized hopscotch map implementations and found to yield significant speedups compared to a server-grade CPU: $17.1x/3.2x$ (area-optimized) and $49.1x/9.2x$ (performance optimized) over standard C$++$ map and hyper-optimized hopscotch map implementations respectively.  The in-situ processing with PIM also minimizes energy spent on moving data, thus achieving energy savings in addition to the performance benefit. Quantifying the energy savings is left for future work. 

\section{HASHMEM ARCHITECTURE}\label{sec:architecture}

{We have designed and implemented a PIM architecture that leverages the lower access latency and inherent parallelism available within the DRAM structure at the subarray level to perform hashmap lookups at the subarray interface. Our key idea behind the design involves mapping an entire hash bucket to a subarray row within PIM memory. Each subarray row contains  between 512-2048 {\em columns}, where a column is defined in terms of a multi-bit access length, with each such column being 4, 8 or 16 bits in length. Internally each of these subarrays could be broken down into mats for implementation purposes. However, we consider a subarray to be a set of mats that are activated in parallel for our understanding of the rest of the paper.

Since an entire hash bucket happens to be mapped to a subarray row, the entire bucket is activated into the row buffer when the subarray row is accessed. We designed processing elements (PEs) to sit at the edge of each subarray closer to the row buffer to perform lookups for necassary keys within the activated hash bucket. Each PE consists of (i) \textbf{comparison unit} to perform the comparison operation, (ii) \textbf{control unit or logic} to orchestrate and control the operations and (iii) \textbf{output register} to hold the resultant value associated with a matched key during the lookup operation. 
In our architecture, we propose two implementations of these comparison units, an {\em area-optimized} version and a {\em performance-optimized} version.

\begin{figure}[h]
  \centering
  \includegraphics[width=\linewidth]{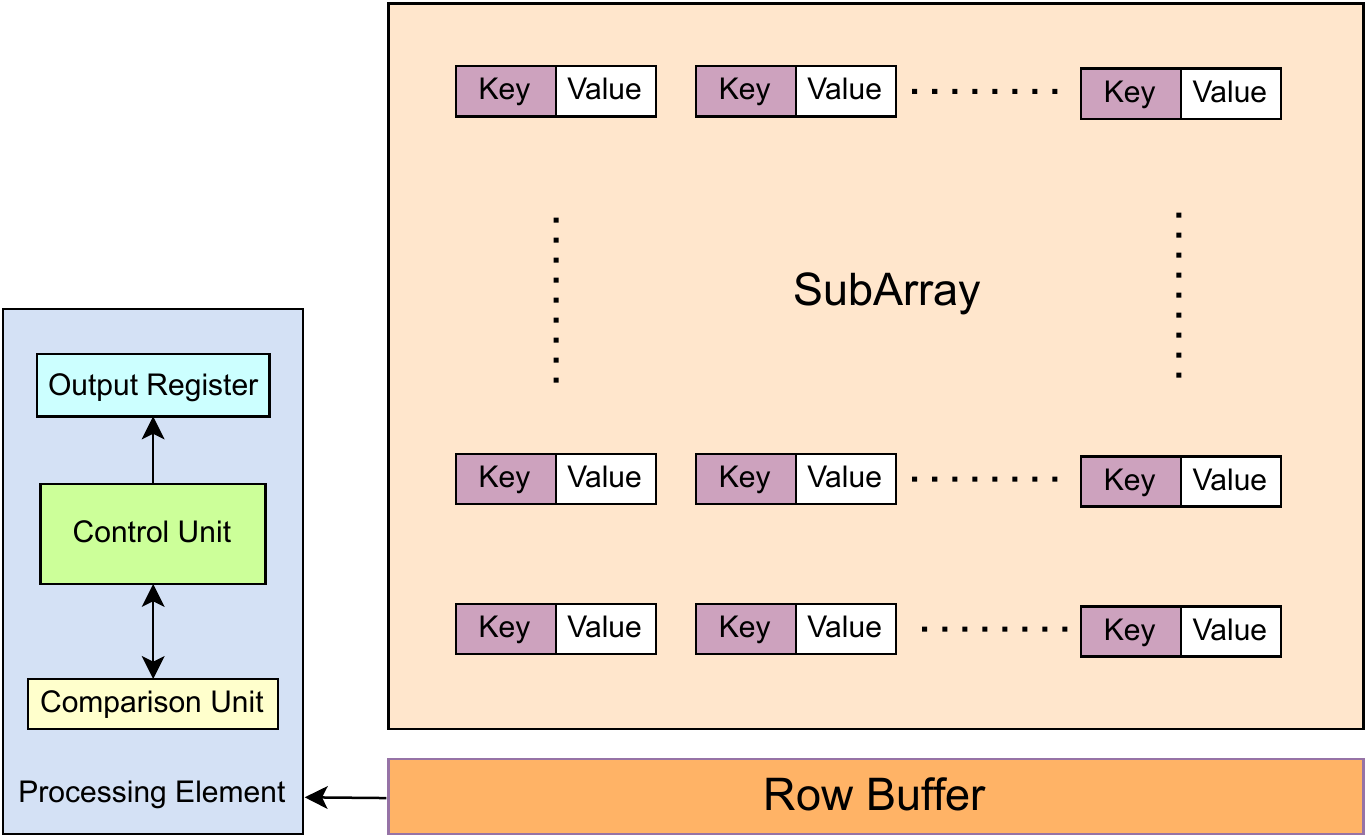}
  
  \caption{Area-optimized HashMem Architecture}
   \label{fig:areaoptarch}
\end{figure}

\subsection{Area-optimized version}

The area-optimized version of HashMem performs the hash bucket traversal in an element-serial bit-parallel manner. Fig. \ref{fig:areaoptarch} demonstrates the architecture for the area-optimized version. The PE accesses each of the activated hash bucket's key-value pairs sequentially from the row buffer and looks for a key match. Upon a match, the corresponding value would be stored within its \textbf{output register} to be read out later by the RLU.

\subsection{Performance-optimized version}
As shown in Fig. \ref{fig:perfoptarch}, the performance-optimized version relies on placing many small comparison units below the row buffer to scan all the keys in parallel within a single or small number of clock ticks. Compared to the Area-optimized version, this offers higher performance with lower execution time but suffers from increased area overhead since many comparison units must be placed and pitch-mapped along the subarray row buffer. This version  resembles the operation of Content Addressable Memory (CAM) except that this is operating on a DRAM row buffer at the subarray-level.

\begin{figure}[h]
  \centering
  \includegraphics[width=0.8\linewidth,keepaspectratio]{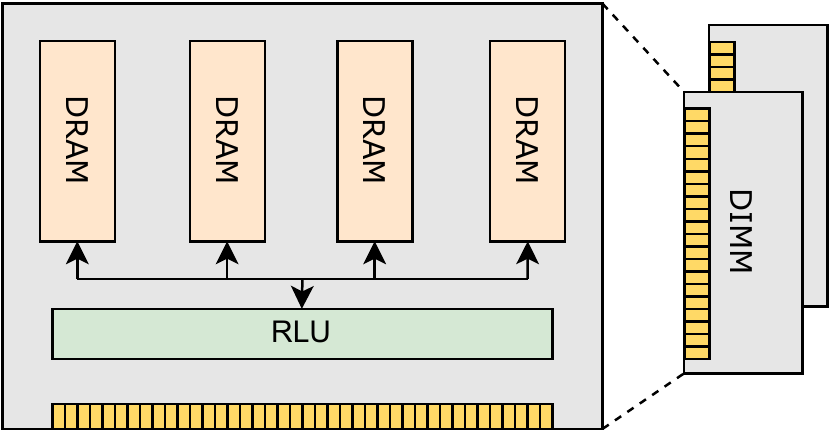}
  \caption{RLU mounted at rank-level}
   \label{fig:rlu}
\end{figure}

\subsection{Rank-Level Unit (RLU)}

The HashMem architecture also involves the usage of a {\em Rank-level Unit (RLU)}, which acts as an intermediary orchestrating agent or command processor between the in-situ PIM processing elements and the host processor (CPU). The job of the RLU is to : 
(i) Propagate the key to be searched to the necessary subarray
(ii) Orchestrate probing operations compliant with the DRAM timing parameters and architecture constraints
(iii) Retrieve the output values after the probing operation is completed from the subarray units and buffer them before transferring them to the memory controller.

The RLU helps in the overall integration with the rest of the system by abstracting the PIM operations and interfacing with the Memory Controller (MC) with special PIM-capable extended DRAM commands. This ensures the host can support the PIM capabilities with minimal changes to its integrated memory controllers. The RLU is responsible for communicating with the in-situ subarray level PIM elements and orchestrating operations amongst them. It is analogous to the command processor (CP) that exists within  GPUs to interface with its Streaming Multiprocessors (SMs) or Compute Units (CUs) and the PCIe bus. As shown in figure \ref{fig:rlu}, since the RLU is mounted as a separate chip, the logic area overhead of the RLU does not affect the memory capacity of the DIMM, as observed in similar rank-level modifications done with AxDIMM \cite{AxDIMM}.

\begin{figure}[h]
  \centering
  \includegraphics[width=\linewidth]{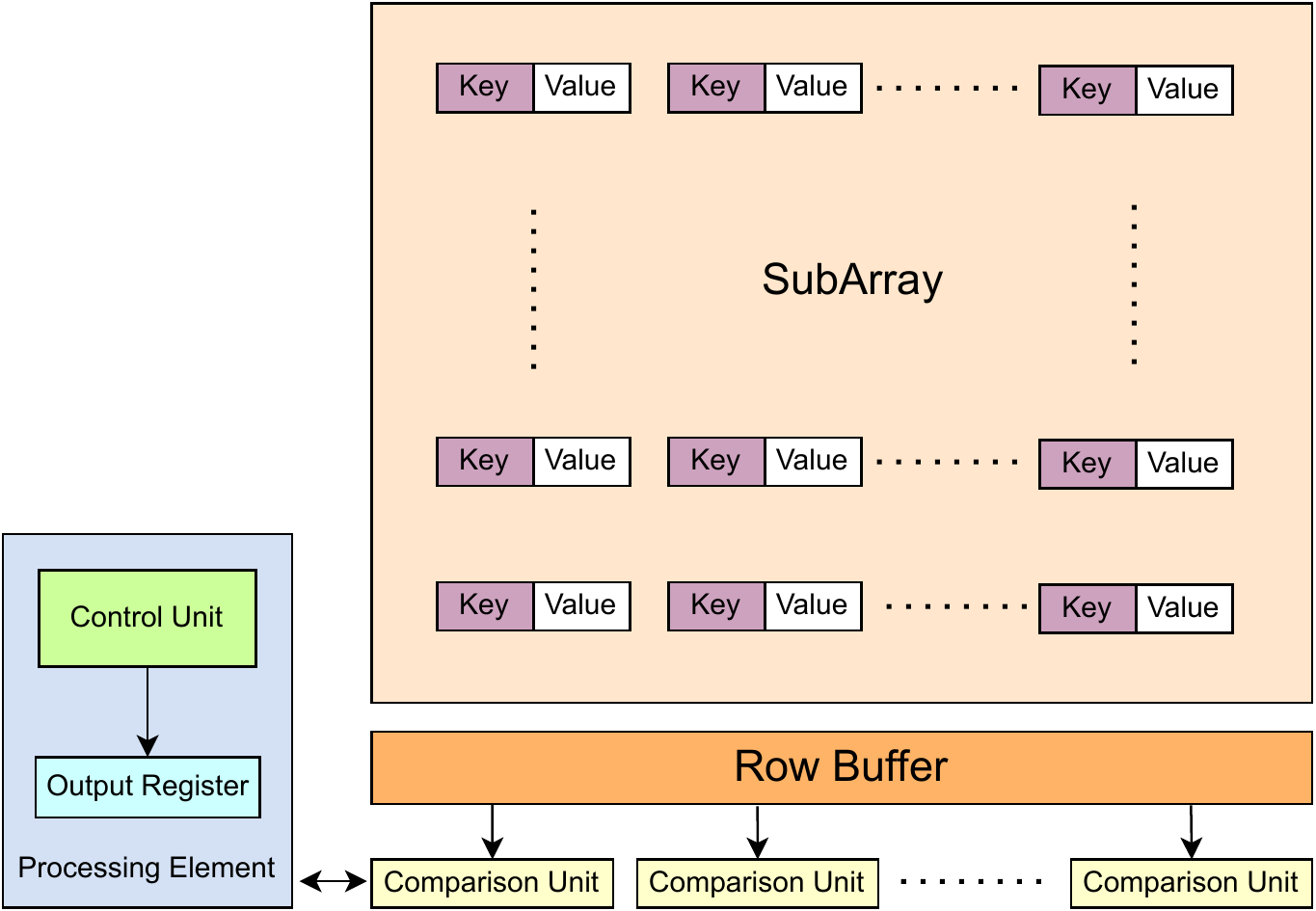}
  \caption{Performance-optimized HashMem Architecture}
   \label{fig:perfoptarch}
\end{figure}

\subsection{Virtualization}

The CPU operates on physical addresses in the typical fashion. Hence, in order to make the PIM memory systems compatible with the current virtualization scheme, we rely on storing hash buckets at page granularity. With this, irrespective of the location of page within physical memory, when a hash bucket is accessed, the corresponding page containing the bucket is activated and its related subarray processing elements are enabled to perform the lookup operations. In scenarios where a page is co-located with other pages in the same row buffer, the page start and end addresses are communicated to the subarray PEs to access only the necessary address range. Using this scheme, the traditional virtualisation is still supported without the need for CPU to concern itself with physical data placement to orchestrate PIM operations.

\subsection{Probing}


Once the initial dataset is populated within the PIM memory, the CPU communicates the key to be probed in the respective hash bucket. The page table translation helps in locating the necessary rank (RLU) and subarray row that holds the hash bucket. The respective RLU receives a compute-capable DRAM command that informs it of the input key to be probed and the address of the page to be probed. The RLU, in turn, orchestrates the probing operation by activating the necessary subarray row and communicating to the in-situ subarray PE to perform the actual probing operation. The RLU later retrieves the value from the output register of the same subarray-level PE and passes it to the Memory Controller (MC) in a cache line format. The cache line can be padded with additional zeroes if the data being transferred is less than the size of a cache line. The MC, after receiving the data from the RLU, places it into the requested CPU's Last Level Cache (LLC) address. Once the CPU reads and extracts the value from the cache, the probing operation is complete.

Currently, deletion operations involve putting tombstone values at the place where a key-value pair is deleted, at the cost of wasted space. This is similar to software implementation of hashmaps. We aim to further investigate how to perform efficient deletion operations to reclaim the space back for further usage while using the hashmap on PIM.

Often times, the load distribution of key-value pairs amongst the hash buckets is not equal and this might lead to some buckets having too many key-value pairs and some having too few. We ran a test case scenario where we mapped the first 350,000 words of a dictionary into a hashmap and measured the length of each bucket. We observed a significant variance in the lengths as shown in fig. \ref{fig:hashchainlen} that demonstrated the under and over-utilisation of buckets. 


\textbf{Under-utilized buckets} - If the page size is N bytes of memory, and the bucket occupies only P bytes of data (where P < N), then the remaining (N-P) bytes of the page are wasted and lead to inefficient memory usage. The page size N is dictated at the boot time without any prior knowledge of the dataset. The value of P (bucket size) is decided during the runtime and depends on the input dataset. Hence, there are bound to be certain buckets which are under-utilized and efforts could be made to map and fit two or more of such buckets into the same page. This helps in reorganising the memory and improving its utilization. However, care has to be taken to ensure proper bookkeeping of the relocated buckets. Also, a strict criterion that the bucket to be relocated is not split and fragmented across multiple pages is to be followed.


\textbf{Over-utilized Buckets} - Some hash buckets may exceed the allocated page size of N bytes and occupy P bytes (where P > N), resulting in an overflow situation that needs to accommodate an extra (P-N) bytes. In these scenarios, an extra page is allocated to accommodate the overflow data, and a bookkeeping structure is updated to record the presence of hash bucket across two or more pages which helps while performing a lookup. Essentially, having these extra pages spread across different channels and ranks helps in probing them in parallel, thereby improving the performance. This optimization could be introduced into the Memory Management Unit (MMU) to instruct it to spread pages containing overly-utilized buckets across different channels evenly to enable the parallel probing of pages. We have marked this as an avenue for future work with micro-architectural changes to be investigated to introduce to support for this optimization strategy.

Alternately, several works such as \cite{perfect-hashing, openadd}, propose ideal hashing functions to counter this unequal distribution phenomenon when certain prior knowledge of the dataset is available to us.  

\begin{figure}[h]
  \centering
  \includegraphics[width=\linewidth]{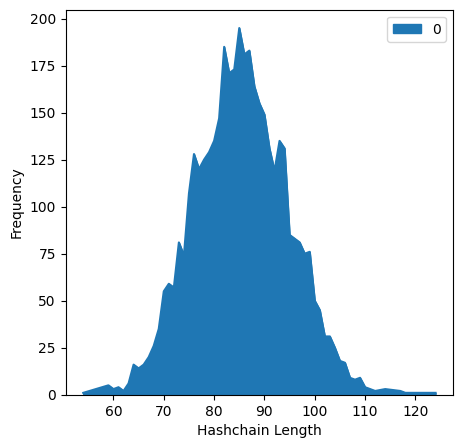}
  \caption{ Length of Hashbuckets after mapping first 350,000 words in a dictionary}
   \label{fig:hashchainlen}
\end{figure}

\subsection{Memory Controller (MC) Changes}

The Memory Controller needs to have the capability to differentiate between a conventional READ/WRITE operation and PIM operation. Specifically, it needs to understand that a particular page being accessed is a hash bucket that needs traversal/lookup to occur on the PIM side and retrieve only the value associated with a specific key. This interaction with the MC should be invoked by the CPU and abstracted from the programmer and exposed as a simple library call. Towards this end, it could be suggested as an extension to the ISA and change on the micro-architectural implementation. Furthermore, we also suggest that the MC have capabilities to communicate with RLU on the other side of the memory bus using special physical layer (PHY) commands that are pin-compatible with the existing DDR standards. The RLU present on the DIMM provides the necessary abstraction to the memory controllers to understand and parse these special PIM commands and orchestrate the operations with its related in-situ subarray-level elements.
\section{Programming Interfaces}

We have provided snippets of code for insertion and lookup in the HashMem PIM-capable memory below. A pseudo-code is listed that provides the necessary abstraction for the programmer to utilize this PIM-capable memory using a high-level programming language. We make use of a bookkeeping structure that keeps track of the hash buckets and the pages that store them. 

\subsection{Insertion Operation}

The overall idea behind the insertion operation is to obtain the hash bucket (or page/pages the bucket is mapped to) that needs to store the key-value pair by performing a hash function on the key. However, there are several constraints, as mentioned in Section \ref{sec:architecture}, that need to be looked after, especially with regard to bucket utilization and page overflow scenarios. 

Initially, in Step-\ding{202}, we obtain the page size from the system information using a library call. This information is usually stored in the operating system after the boot procedure has initialized the page tables and other virtualization structures. In \ding{203}, we hash the key and decide the bucket and page in which to store the key-value pair. Consequently, we need to perform a check if the bucket is going to overflow in step-\ding{204} while inserting the input-key-value pair. If the bucket is not going to overflow, then the key-value pair could be inserted successfully, as in Step-\ding{205}. However, if the bucket is about to overflow, we do pim\_malloc() to initialize a fresh page in step-\ding{206}. We update a bookkeeping structure to reflect that a particular hash bucket extends to a new page. This is to ensure any subsequent lookup operations probe the new page in addition to the prior existing page/(s). Once the bookkeeping structure is updated, we go ahead and store the key-value pair on the new page (step-\ding{207}) and provide the necessary return code to reflect the successful insertion of the key-value pair.

\lstset{frame=tb,
  language=c++,
  aboveskip=5mm,
  belowskip=5mm,
  escapechar=<>,
  showstringspaces=false,
  columns=flexible,
  basicstyle={\small\ttfamily},
  numbers=none,
  numberstyle=\tiny\color{blue},
  keywordstyle=\color{red},
  commentstyle=\color{pink},
  stringstyle=\color{green},
  breaklines=true,
  breakatwhitespace=true,
  emph={pageSizeInBytes, getPageSizeFromSystemInfo, numOfKVPairsPerPage, destinationPage,
        getHashValueFromHashingAlgorithm, currentNumOfKVPairsInPage, getCurrentPageSize
        currentSizeOfPage, storeKVPairIntoPage, inputKey, inputValue, pim_page, newPage,
        ret_code, PP_ERROR, updateBookkeepingStructure, PR_SUCCESS, pim_malloc, currentSizeOfPage},
    emphstyle={\color{blue}},
  commentstyle=\color{commentgreen},
  tabsize=4}

\begin{lstlisting}[caption=PIM Insertion Operation]
    void MapInputKeyValuePairToHashMemPage(keyDataType inputKey, valueDataType inputValue) {

	pageSizeInBytes = getPageSizeFromSystemInfo(); <>\ding{202}<>
			/* This Page Size is decided during the boot time based on subarray structures. This information is retrieved using the runtime system information call. */

	numOfKVPairsPerPage = pageSizeInBytes / sizeOfEachKVPair;	
            // Calculating number of KV pairs per page
	
	destinationPage = getHashValueFromHashingAlgorithm(inputkey);	<>\ding{203}<>
			// Perform Hashing operation to decide which page the key-value pair needs to be mapped to

	currentNumOfKVPairsInPage = getCurrentPageSize(destinationPage);
			// Get size current page size in terms of how many number of key-value pairs it is currently holding

	/* Perform a check if the page can accomodate the key-value pair or not */
	if(currentSizeOfPage < numOfKVPairsPerPage) <>\ding{204}<> { 
		storeKVPairIntoPage(destinationPage, inputKey, inputValue);	<>\ding{205}<>
        // Store the KV pair into the destinationPage
	}
	else{  // if page is already full
		pim_page	newPage;
		ret_code = pim_malloc(newPage);		<>\ding{206}<>
        // allocates a new page and assigns to newPage structure
		if(ret_code == PR_ERROR)
			return PR_ERROR;		// Page allocation failed, function exits

		updateBookkeepingStructure(newPage, destinationPage); 	<>\ding{207}<>	
				/* Attaches and links new page to old page (destination_page) in a Linked List fashion. This book-keeping structure also informs that these two pages (new page and old page) need to be probed together when looking for keys. Because the key can reside either in new page or old page. */
		storeKVPairIntoPage(newPage, inputKey, inputValue);
				// Store the key-value into the newPage
	}
	return PR_SUCCESS;
}

\end{lstlisting}

\subsection{Lookup operation}

The lookup operation is relatively simple and straightforward, where the user provides an input key and expects the value associated with it to be returned. In Step-\ding{202}, the hashing function computes the bucket and page to be probed based on the input key provided. In Step-\ding{203}, we make use of a special library call that performs a hash bucket lookup operation. In the background process, the library call consults the bookkeeping structure to check the number of pages to probe and instructs the MMU to perform page-level probing operations that traverses the entire hash bucket using the subarray PIM processing elements. These PIM probing operations retrieve the value associated with the key and provide the necessary return code to reflect the successful lookup operation.

\lstset{frame=tb,
  language=c++,
  aboveskip=5mm,
  belowskip=5mm,
  escapechar=<>,
  showstringspaces=false,
  columns=flexible,
  basicstyle={\small\ttfamily},
  numbers=none,
  numberstyle=\tiny\color{blue},
  keywordstyle=\color{red},
  commentstyle=\color{pink},
  stringstyle=\color{green},
  breaklines=true,
  breakatwhitespace=true,
  emph={bucketToProbe, getHashValueFromHashingAlgorithm, valueDataType, inputKey, outputValue, pimProbeBucket, 
        outputValue, PR_ERROR},
    emphstyle={\color{blue}},
  commentstyle=\color{commentgreen},
  tabsize=4}
\begin{lstlisting}[caption=PIM Probe Operation]
    <>{\color{blue}<>
    valueDataType probeKey(keyDataType inputKey) {

	bucketToProbe = getHashValueFromHashingAlgorithm(inputkey);    <>\ding{202}<>
		// Perform hashing to get the hash bucket to probe

    /* Variable to hold value associated with key-value pair */
	valueDataType outputValue = NULL;

	outputValue = pimProbeBucket(bucketToProbe);   <>\ding{203}<>
			/* This first checks the book keeping structure as to how many pages to probe and then issues a special pim probe command to the memory controller to perform probing on PIM-capable memory*/

	/* If the value was not found */
	if (outputValue == NULL)
		return PR_ERROR;

	return outputValue;
}
\end{lstlisting}}
\section{Evaluation}

In this section, we will discuss the performance and area overhead of HashMem architecture and compare it with traditional CPU-based implementation baselines. The configuration of the hardware setups utilized in our analysis are as mentioned in the Table-\ref{tab:table_config_details}.

\begin{table}[h!]
\scriptsize
\centering
\caption{\centering{Hardware Configuration}\vspace*{-7pt}} 
\begin{tabular}{@{}ll@{}}
\toprule
Property                  & Value                              \\ \midrule
Processor Name            & Intel(R) Xeon(R) Silver 4208 CPU (2.1 GHz)\\
Total Cores / Main Memory              & 8  (16 threads) / 512 GB                              \\
L1D/L2/L3 Cache Size & 32 KB per core/1 MB per core/11.2 MB shared                             \\ \midrule
                            
HashMem                 &   DDR4\_8Gb\_x16\_3200  (Single Channel)  \\
                        &   8 banks per rank, 128 subarrays per bank  \\
                        &   512 rows per subarray   \\
                            \bottomrule
\end{tabular}
\label{tab:table_config_details}
\end{table}

\begin{table}[h!]
\scriptsize
\centering
\caption{\centering{Workload Overview}\vspace*{-7pt}} 
\begin{tabular}{@{}ll@{}}
\toprule
Property                  & Value                              \\ \midrule
Dataset                & Contains 100 million key-value pairs (800 MB) \\
                        & 10\% i.e. 10 million randomly selected keys probed \\
                        \midrule
Points of Comparison    & Standard C++ Map (binary tree)    \\
                        & C++ unordered\_map (hashmap)  \\
                        & hopscotch map (optmized hashmap)  \\
                                                
                            
                            \bottomrule
\end{tabular}
\label{tab:software_configuration}
\end{table}

\subsection{Benchmarking}

Currently, there exists no standard benchmark to test exclusively for the hashmap performance. Hence, we proposed our microbenchmark to test the hashmap performances on both the CPU and HashMem. The details of the microbenchmark and baselines we compare against are provided in the proceeding section. In order to estimate and model the HashMem performance, we analyzed the timing data gathered from prior works \cite{dont-forget-io-paper, i7-performance-tuning-guide, mem-systems-jacobs, dram-sim3}.

\subsubsection{Microbenchmark} \label{sec:microbenchmark} 
One of the goals of the microbenchmark is for the input dataset to be sufficiently large enough such that it flows out of the cache and into the DRAM. This ensures that the cache effects of the CPU are sufficiently eliminated and the expensive off-chip DRAM accesses are captured while performing lookups during microbenchmark run. Another goal is to generate random accesses, i.e., accessing random keys to eliminate any prefetching based on spatial locality. 

Towards this end, we propose a microbenchmark consisting of 100 million key-value (KV) pairs with a key and value occupying $4$ bytes each. The key and the value are coded as default uint32\_t data type in C$++$. Hence, each KV pair would occupy $8$ bytes of data, and the overall memory footprint of the input dataset would be $800$ MB in size, sufficiently large enough to overflow the L3 cache of most processors used in our testing scenario. Furthermore, $10\%$ of the keys in the input dataset are probed, which equates to $10$ million keys being searched for in the hashmap. The keys to be probed again are selected at random and fed into both the CPU and PIM architectures simultaneously to assess the hashmap probing performance. 

Although this benchmark does not account for  string values and other types of data, we envision that they could be pre-processed and dictionary-encoded into numerical values to be used in HashMem. Performing string value comparison or any regex operations at the subarray-level units would incur a very high area overhead and is avoided. Hence we consciously supported probing just numerical values with HashMem PIM architecture. 

\subsection{Performance of different data structures on CPUs}

\begin{figure}[h]
  \centering
  \includegraphics[width=\linewidth]{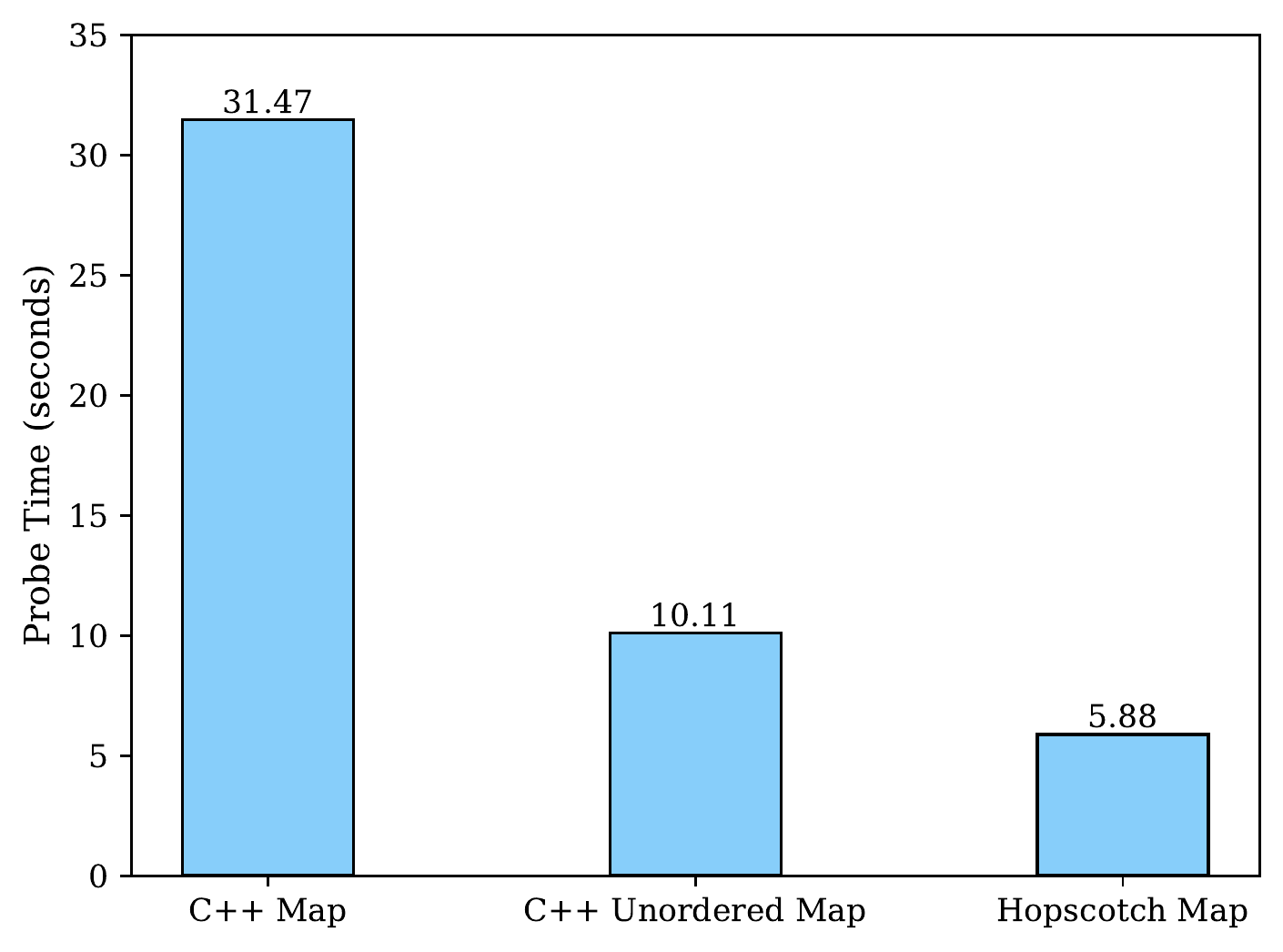}
  \caption{Probing Times of Different Data Structures}
   \label{fig:probingtimecpu}
\end{figure}

Our initial research indicated that “unordered\_map” Standard Template Library (STL) in C++ has the closest implementation resembling a hashmap. However, there are several other default C++ data structures that perform similar operations as a hashmap but are implemented using alternative techniques. One of the most popular examples is the “map” data structure that is implemented as a red-black tree, a specialized binary search tree data structure. We intended compare against both the unordered\_map and map to assess the performance. 

Apart from the default C++ libraries, there are various implementations that are more performance-optimized to yield better throughput in performing key-value pair lookups. Hopscotch is one example that implements a hopscotch hashing algorithm \cite{hopscotch-hashing-paper} to resolve hashing collisions \cite{hopscotchhashing}. We found a popularly used repository online \cite{hopscotch} that implemented this hashing mechanism that we could consider as a state-of-the-art baseline. Interestingly, we discovered that hopscotch\_map is faster than Google’s sparse-hash \cite{sparse-hash} implementation during our testing. We intended our evaluations to compare with a mix of candidates, hence our choice of these three  software implementations. 


The performance results of the three data structures chosen on CPU are demonstrated above in Figure-\ref{fig:probingtimecpu}. Hopscotch is significantly faster than the other implementations by a wide margin, around $5.3x$ and $3.1x$ compared to C$++$ map and unordered\_map, respectively. The map structure performs the worst considering that it is implemented as a balanced binary search tree with many indirect accesses to traverse along the tree both during insertions and probing. The complexity of a map is  $log_2 ^{n}$. Interestingly, we found that unordered\_map rehashes itself when the number of elements to be inserted exceeds the load factor, i.e., the number of buckets available, rather than having more nodes attached to extend each bucket size. 

\subsection{Area overhead}
To estimate the area overhead of subarray-level PIM processing elements, we first obtain the area breakdown of the DDR4 chip (DDR4\_8Gb\_x8\_3200) that is used to build the HashMem. Each subarray contains 512 rows, and there are 128 subarrays per bank. The comparator units of HashMem are implemented in RTL and the delay, area overheads are evaluated using Synopsys DC Compiler in 14nm. We use scaling factors from \cite{cmosscaling} to scale the results  to 22nm. 

\subsubsection{Performance optimized}
For the performance-optimized version, there is a requirement to pitch-map the comparator units to fit within the section boundaries of the row buffer containing the key-value pair segments. This presents significant challenges related to the evaluation of this version and this is part of the future work to investigate HashMem further. 

\subsubsection{Area optimized}
The area-optimized version does not require significant efforts to pitch-map as was required with the performance-oriented version. Our estimate revealed that incorporating 64 additional ALUs, shared by 128 subarrays per bank, incurred only 5.26\% area overhead.

\section{Results}


The performance of hashmap workloads are both dependent on the size and the distribution of the input dataset. The numbers reported are for the evaluation of the dataset as detailed in section \ref{sec:microbenchmark} describing the microbenchmark.

\begin{figure}[h]
  \centering
  \includegraphics[width=\linewidth]{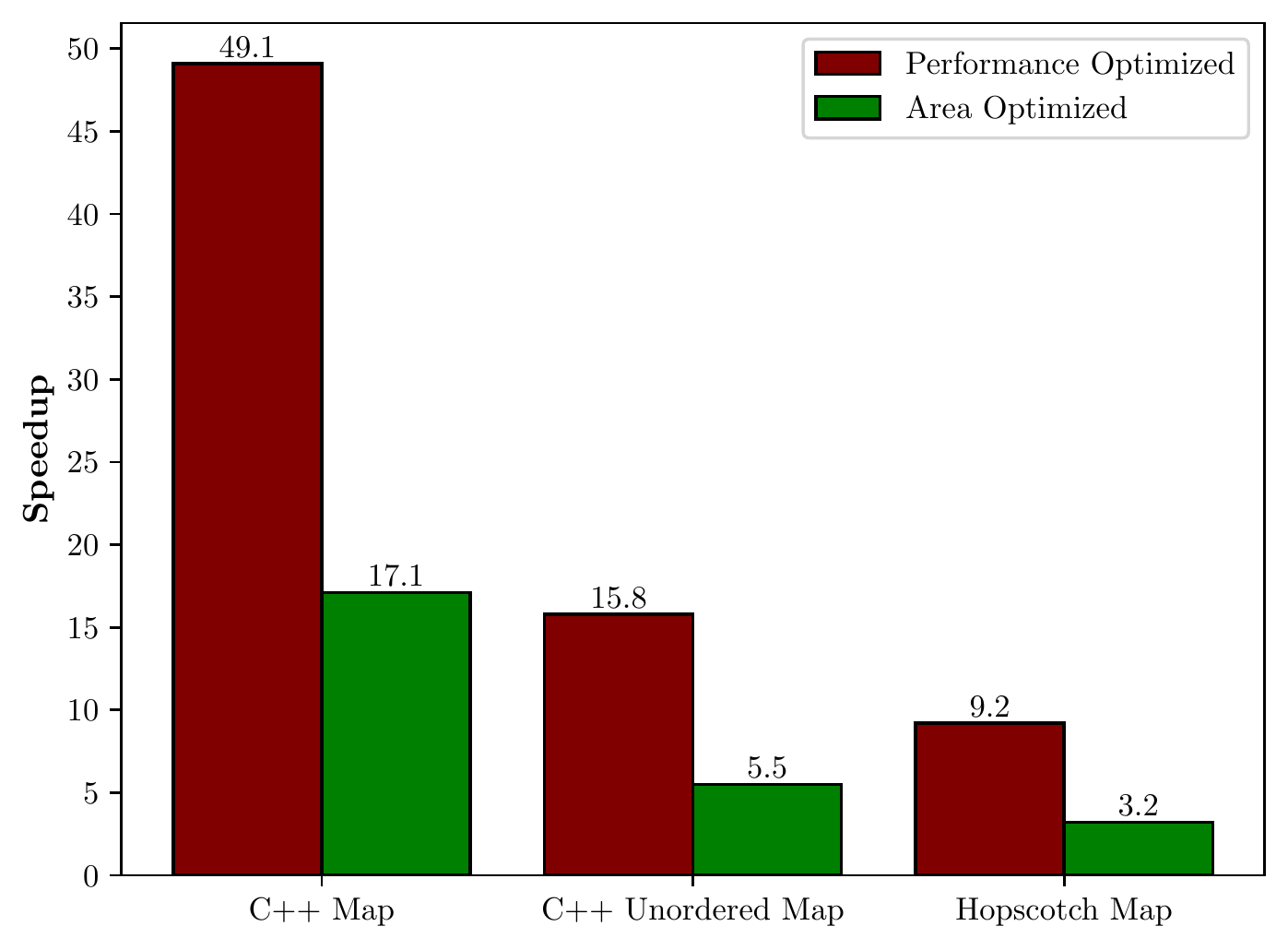}
 \caption{HashMem Speedup Against CPU}
  \label{fig:pimspeedup}
\end{figure}

Figure \ref{fig:pimspeedup} highlights that our area-optimized HashMem version outperforms the standard map, unordered map, and hopscotch map implementations, achieving speedup values of approximately $17.1x$, $5.5x$, and $3.2x$ respectively.

As shown in figure \ref{fig:pimspeedup}, our performance-optimized HashMem surpasses the standard map, unordered map, and hopscotch implementations by even greater factors of $49.1x$, $15.8x$, and $9.2x$ respectively.

In both scenarios, HashMem outperforms the CPU baseline by a wide margin, even against the state-of-the-art hopscotch implementation. An interesting aspect of the experiment is that these results represent a single DRAM channel competing against a server-class CPU. We could parallelize the lookups across the independent memory channels and obtain further improvement in the HashMem performance. These results  demonstrate the potential of PIM architectures, which have tremendous intrinsic parallelism and bandwidth that are not being harnessed  by current Von Neumann-bottlenecked architectures. 

\section{FUTURE WORK}

Our future work broadly aims at improving the architecture  with a richer set of evaluations. We aim to look at a wider variety of datasets with different distribution patterns of key and value strings to assess the hashmap performance in each setting. 
 
\textbf{Tiered-Latency DRAM.} \cite{tiered-latency} splits the subarray into low-latency and higher-latency regions by placing isolation transistors to alter the DRAM READ access timings. We aim to leverage this work to map our hashmap buckets into the low-latency region for faster lookups. In a conventional architecture, mapping several elements to a smaller subset of buckets could degrade the performance. However, due to parallelism provided by the PIM processing elements and the lower-latency offered by the Tiered-Latency DRAM, it could improve our performance significantly. Moreover, the lower the number of buckets available within a hashmap, the higher the probability of a row hit which further leads to improved performance and decreased latency. 

\textbf{Channel-level Parallelism.} Memory channels are independent and parallel READ / WRITE operations could be performed on each of the channels separately. Hence, if there are multiple keys to be probed, these probing operations could be parallelised amongst different channels to increase the throughput. However, this type of parallelism could be exploited only if the keys being probed belong to different channels.

\textbf{Energy Savings Analysis.} Due to the vastly reduced number of expensive off-chip data access over the memory bus, there are significant energy savings to be realised. This also leads to reduced instruction overhead on the CPU related to performing a scan operation on the hash bucket traversals.

\textbf{Data Types}. Currently our evaluations only support 32-bit int operations on both key and value. The limitation is with regards to the configuration of the subarray-level PIM elements. We aim to investigate re-configurable sizes for future evaluations to support other data types too. 

\textbf{Hash Function.} We aim to investigate an optimum hashing mechanism that can evenly distribute the input dataset over several buckets of equal or near equal length. This is to reduce certain buckets from getting over-utilized and some from getting under-utilized. There are prior works in this area such as \cite{perfect-hashing}.

\textbf{Real-world kernels.} Many genomics, database and other applications extensively make use of hashmap structures in their kernels while implementing them. We aim to test our PIM architecture in these settings and observe the performance improvements at the application level. 

\section{Conclusions}

We proposed HashMem, a subarray-level PIM architecture that accelerates hashmap lookups by leveraging the existing parallelism available within the DRAM structure. We have demonstrated that the two variants of HashMem, area and performance-optimized ones, were able to outperform state-of-art hopscotch hashmap and default C++ map implementations on CPU by at least $3.2x/9.2x$ and $17.1x/49.1x$ respectively. The area overhead as observed on the area-optimized version was found to be $5.26\%$. We have laid out several directions as part of our future work to further investigate potential of HashMem architecture.

\section{Acknowledgements}

This work was supported in part by PRISM, one of seven centers in JUMP 2.0, a Semiconductor Research Corporation (SRC) program sponsored by DARPA.  We also thank the reviewers for their helpful feedback and suggestions.


\bibliographystyle{IEEEtranS}
\bibliography{refs}

\end{document}